\documentclass[journal]{IEEEtran}

\usepackage{amsmath}
\usepackage{amssymb}
\usepackage{amsthm}
\usepackage{graphicx}
\usepackage{epstopdf}
\usepackage{bm}
\usepackage{color}
\usepackage{cite}
\usepackage{subcaption}
\usepackage{dsfont}
\usepackage{algorithm}
\usepackage{algpseudocode}
\usepackage{enumitem}

\newtheorem{theorem}{Theorem}

\newtheorem{assumption}{Assumption}
\newtheorem{remark}[theorem]{Remark}

\begin{document}
\title{A Logistic Regression Approach to Field Estimation Using Binary Measurements} 

\author{Alex S.~Leong, Mohammad Zamani, and Iman Shames
\thanks{A.~Leong and M.~Zamani are with Land Division, Defence Science and Technology Group, Melbourne, Australia, email \texttt{\{alex.leong, mohammad.zamani\}@defence.gov.au }. I.~Shames is with the CIICADA Lab, College of Engineering and Computer Science, Australian National University, Canberra, Australia, e-mail: \texttt{iman.shames@anu.edu.au}  }}

\maketitle

\begin{abstract}
In this letter, we consider the problem of field estimation using binary measurements. Previous work has formulated the problem as a parameter estimation problem, with the parameter estimation carried out in an online manner using sequential Monte Carlo techniques. In the current work, we consider an alternative approach to the parameter estimation based on online logistic regression. The developed algorithm is less computationally intensive than the sequential Monte Carlo approach, while having more reliable estimation performance. 
\end{abstract}

\section{Introduction}
The mapping of a contaminated area due to a chemical, biological, radiological, or nuclear (CBRN) leakage is an important task in allowing operations to be carried out in the affected area. The use of mobile autonomous vehicles in carrying out this mapping is likely to be crucial, as existing infrastructure for detecting/sensing such contaminants in the affected area is often lacking.  

Previous work in field estimation using mobile autonomous vehicles include \cite{LaShengChen,RazakSukumarChung_journal,LeongZamani_SP}. The field was modelled as a sum of radial basis functions, with the field estimation problem then reduced to estimating the parameters of the basis functions. 

Motivated by the limitations of current technologies in chemical sensors \cite{RosserPavey_remote_sensing,RisticGunatilakaGailis_IF}, in \cite{LeongZamani_SP} it was assumed that the noisy sensor measurements were coarsely quantized, and in particular binary \cite{RisticGunatilakaGailis,Selvaratnam_CDC}. The field parameters were estimated using sequential Monte Carlo techniques \cite{Doucet_book}, and an active sensing mechanism \cite{Kreucher_proceedings,RisticMorelandeGunatilaka} for choosing the next measurement location given the measurements currently collected, was proposed. 

In this letter, we also consider the use of binary measurements. Instead of the sequential Monte Carlo approach, we instead propose an approach that views the parameter estimation problem as an online logistic regression problem \cite{JezequelGaillardDuri}. The parameters are updated recursively using a Newton-type update \cite{LesageLandryTaylorShames}. With further approximations, the computational load can be significantly reduced from that of the sequential Monte Carlo approach. The active sensing mechanism of \cite{LeongZamani_SP} based on maximization of an information measure does not extend in a straightforward way to the current approach, and thus an alternative active sensing mechanism is proposed based on numerical properties of the Hessian of the cost function. 

The letter is organized as follows. Section \ref{sec:system_model} gives the system model. The logistic regression approach is presented in Section \ref{sec:logistic_regression}, first in an ``exact'' form using online Newton's method (ONM). Section \ref{sec:approximate_Newton} then introduces an approximate ONM that is computationally less intensive than the exact counterpart. An active sensing mechanism, inspired by the results of the preceding sections, is presented in Section \ref{sec:active_sensing}.  Simulation results that illustrates the performance of the proposed methods are presented in Section \ref{sec:numerical}.  

\emph{Notation}: The operators $\sigma_{\textnormal{min}}(\cdot) $ and $\lambda_{\textnormal{min}}(\cdot)$ denote the minimum singular value and minimum eigenvalue of their arguments, respectively.

\section{System Model}
\label{sec:system_model}
We assume that the field can be sufficiently well approximated \cite{ParkSandberg} by
\begin{equation}
\label{field_model}
\phi(\mathbf{x}) = \sum_{i=1}^p \beta_i K_i(\mathbf{x})
\end{equation}
where $\mathbf{x} \in \mathbb{R}^2$ is the position, $K_i(\cdot)$ are basis functions, and $\beta_i$ are the  coefficients multiplying the basis functions. Similar models have been used in e.g. \cite{LaShengChen,MorelandeSkvortsov,RazakSukumarChung_journal,LeongZamani_SP}. For the basis functions we will use the Gaussian kernel
$$K_i(\mathbf{x}) = \exp \left(- \frac{\|\mathbf{c}_i-\mathbf{x}\|^2}{\sigma_i^2}\right)$$
where $\mathbf{c}_i \in \mathbb{R}^2 $ represents the center of the $i$-th basis function. Note that this model captures a pollutant field at steady-state in a sufficiently rich, yet tractable analytic model. In reality the physical processes that generate such a field may be a lot more complicated. However, this model serves as a surrogate that approximates the true field provided that a sufficient number of basis functions are considered \cite{ParkSandberg}. 

For the measurement model, at position $\mathbf{x}$, the sensor is first assumed to have a noisy measurement
\begin{equation}
\label{eqn:noisy_measurement}
y(\mathbf{x}) = \phi(\mathbf{x}) + v(\mathbf{x}),
\end{equation}
where $v(\mathbf{x}) $ is noise. This noisy measurement is then thresholded to give a binary measurement
$$z(\mathbf{x}) = \mathds{1} (y(\mathbf{x}) > \tau),$$
where $\tau$ is a positive scalar threshold, and $\mathds{1}(\cdot)$ is the indicator function that returns 1 if its argument is true and 0 otherwise. 

In the absence of knowledge of the ``true'' parameters, we assume that $\mathbf{c}_i$ and $ \sigma_i^2, i=1,\dots,p$ are to be chosen by us.\footnote{This corresponds to case (i) of \cite{LeongZamani_SP}. Case (ii), which estimates all $\beta_i$, $\mathbf{c}_i$, $ \sigma_i^2, i=1,\dots,p$, was also considered in \cite{LeongZamani_SP}, but was found to give less reliable estimates.}  The objective is to estimate the parameters $\bm{\beta} \triangleq (\beta_1,\dots,\beta_p)$ from binary measurements collected by an autonomous vehicle, where the vehicle can collect measurements from any position in the area of interest $\mathcal{S}$.

\section{Logistic regression approach}
\label{sec:logistic_regression}
We consider a cost function based on logistic regression. 
Recall the logistic function $\ell(x) = 1/(1 + \exp(-\eta x))$,
where $\eta > 0$ is a parameter. Larger values of $\eta$ will more closely approximate the function 
$\mathds{1}(x > 0)$.

Let 
$\tilde{z} \triangleq 2 z - 1.$ Since $z \in \{0,1\}$, we have $\tilde{z} \in \{-1,+1\}$.
Consider the following cost function, 
$ J_k (\bm{\beta}) = \sum_{t=0}^k g_t (\bm{\beta})$
with per stage cost
$$g_t(\bm{\beta}) = \log \left(1 + \exp(- \eta \tilde{z}_t (\bm{\beta}^T \mathbf{K}(\mathbf{x}_t) - \tau))\right), $$
where $\mathbf{K}(\mathbf{x}) \triangleq \begin{bmatrix}
K_1(\mathbf{x}) & K_2(\mathbf{x}) & \dots & K_p(\mathbf{x})
\end{bmatrix}^T$ and $t$ is a time/iteration index. The measurement $\tilde{z}_t$ and position $\mathbf{x}_t$ are assumed to be provided to us via a sensor and e.g. GPS, respectively. 
The case where $\eta=1$ is similar to the costs used in logistic regression problems, see e.g.  \cite[p.516]{CalafioreElGhaoui}.

We are interested in \textit{online} estimation of $\bm{\beta}$ recursively rather than solving an optimization problem in batch mode. In this setup a new measurement may be collected after each time step. We use the online Newton's method (ONM) of \cite{LesageLandryTaylorShames}.
To this aim, the gradient of $J_k (\bm{\beta}) $ is computed as the following:
\begin{equation}
\label{eqn:grad_J}
\begin{split}
\nabla J_k (\bm{\beta}) & = \sum_{t=0}^k \nabla g_t (\bm{\beta}) \\ & =  \sum_{t=0}^k \frac{ - \eta \tilde{z}_t }{1+\exp(\eta \tilde{z}_t(\bm{\beta}^T \mathbf{K}(\mathbf{x}_t) -  \tau))} \mathbf{K}(\mathbf{x}_t).
\end{split}
\end{equation}
Similarly, the Hessian of $J_k (\bm{\beta}) $ is
\begin{equation} 
\label{eqn:Hessian_J}
\begin{split}
&\nabla^2 J_k (\bm{\beta})  = \sum_{t=0}^k \nabla^2 g_t (\bm{\beta}) \\ & = \sum_{t=0}^k \frac{ \eta^2 \tilde{z}_t^2 \exp(\eta \tilde{z}_t(\bm{\beta}^T \mathbf{K}(\mathbf{x}_t) -  \tau)) }{\big(1+\exp(\eta \tilde{z}_t(\bm{\beta}^T \mathbf{K}(\mathbf{x}_t) -  \tau)) \big)^2}  \mathbf{K}(\mathbf{x}_t) \mathbf{K}^T(\mathbf{x}_t).
\end{split}
\end{equation}
The ONM for recursively estimating $\bm{\beta}$ is then given by
\begin{equation}
\label{eqn:online_Newton_exact}
 \bm{\hat{\beta}}_{k+1} = \bm{\hat{\beta}}_k - \left( \nabla^2 J_k (\bm{\hat{\beta}}_k) \right)^{-1} \nabla J_k (\bm{\hat{\beta}}_k),
\end{equation}
where $ \nabla J_k (\bm{\hat{\beta}}_k) $ and $\nabla^2 J_k (\bm{\hat{\beta}}_k)
 $ can be obtained by substituting  $\bm{\hat{\beta}}_k$ into \eqref{eqn:grad_J} and \eqref{eqn:Hessian_J}, respectively. 

\subsection{Performance of online Newton's method}
\label{sec:convergence}

Define $\bm{\beta}^*_k = \arg\min_{\bm{\beta}} J_k (\bm{\beta})$. We make the following:
\begin{assumption} $~$
\label{assumption_Lesage_Landry}
\begin{enumerate}[label=(\roman*)]
    \item There exists a positive real scalar $\bar{v}$ such that $\|\bm{\beta}^*_{k+1} - \bm{\beta}^*_k\|\leq \bar{v}$ for all $k>0$.
    \item There exist positive real scalars $h_k$ such that $\|(\nabla^2 J_k(\bm{\beta}^*_k))^{-1}\| \leq 1/h_k$.
    \item In a ball of radius $\rho$ centered at $\bm{\beta}^*_k$, the Hessian is Lipschitz continuous with Lipschitz constant $L_k$.
    \item There exist positive integer and real scalars $\overline{k}$ and $\underline{\lambda}$ such that $ \lambda_{\textnormal{min}} \left ( \sum_{t=\tau}^{\overline{k}+\tau-1} \nabla^2 g_t (\bm{\beta})\big) \right ) \geq \underline{\lambda}, \, \forall \tau. $
\end{enumerate}
\end{assumption}
Assumptions (i)-(iii) are assumptions from \cite{LesageLandryTaylorShames}, while assumption (iv) is a ``persistence of excitation'' type assumption that there is sufficient richness in our collected measurements. 

We aim to bound the regret of the ONM defined by
\begin{align}
    \mathrm{Reg}(T) = \sum_{k=0}^T \left(J_k(\hat{\bm{\beta}}_k) - J_k(\bm{\beta}^*_k) \right).
\end{align}


For a bound on $\|(\nabla^2 J_k(\bm{\beta}^*_k))^{-1}\| $, suppose we use the induced 2-norm. Then 
\begin{align*}
\|(\nabla^2 J_k(\bm{\beta}^*_k))^{-1}\|_2 & = \frac{1}{\sigma_{\textnormal{min}} \big( \sum_{t=1}^k \nabla^2 g_t (\bm{\beta}^*_k) \big)} \\ & = \frac{1}{\lambda_{\textnormal{min}} \big( \sum_{t=1}^k \nabla^2 g_t (\bm{\beta}^*_k)\big)}   \leq \dfrac{1}{\lceil k/\overline{k} \rceil \underline{\lambda}},
\end{align*}
where the last inequality holds by Assumption \ref{assumption_Lesage_Landry} (iv) and Corollary 4.3.15 of \cite{HornJohnson}. Thus we can set $h_k \triangleq \lceil k/\overline{k} \rceil  \underline{\lambda}$.

For a Lipschitz constant $L_k$ of the Hessian of $J_k(\cdot)$, we can obtain $L_k \triangleq k L_g$ that increases linearly with $k$, since one has a Lipschitz constant $L_g$ for the Hessian of each $g_t(\cdot)$, and there are $k$ such terms in the summation in $J_k(\cdot)$.

For any iterations such that 
$$\| \bm{\hat{\beta}}_k - \bm{\beta}^*_k\| < \min\{\rho, \frac{2   \underline{\lambda}}{3 \overline{k} L_g} \} \triangleq \gamma,$$
then $\| \bm{\hat{\beta}}_k - \bm{\beta}^*_k\| < \min\{\rho, \frac{2 h_k}{3 L_k} \}$, since
\begin{align*}
\| \bm{\hat{\beta}}_k - \bm{\beta}^*_k\|& < \min\{\rho, \frac{2   \underline{\lambda}}{3 \overline{k} L_g} \} = \min\{\rho, \frac{2 k/\overline{k} \underline{\lambda}}{3 k L_g} \} \\ & \leq  \min\{\rho, \frac{2 \lceil k/\overline{k} \rceil  \underline{\lambda}}{3 k L_g} \} = \min\{\rho, \frac{2 h_k}{3 L_k} \}.
\end{align*}
Assume also that $\bar{v} \leq \gamma - (3 \overline{k} L_g  / 2 \underline{\lambda}) \gamma^2.$ 
Then the regret of ONM satisfies \cite{LesageLandryTaylorShames}:
\begin{equation}
 \label{eqn:regret_online_Newton}
 \mathrm{Reg}(T) \leq \dfrac{\kappa}{1-\dfrac{3\overline{k} L_g}{2 \underline{\lambda}}\gamma} (V_T +\delta)   
\end{equation}
where $\kappa$ is a positive scalar such that if $\| \bm{\beta} - \bm{\beta}^*_k\| \leq \gamma$ then $|J_k(\bm{\beta}) - J_k(\bm{\beta}^*_k)| \leq \kappa \| \bm{\beta} - \bm{\beta}^*_k\|$ for $k=1,\dots, T$. Moreover, $V_T = \sum_{k=1}^T \| \bm{\beta}^*_k - \bm{\beta}^*_{k-1}\|$ and $\delta= (3 \overline{k} L_g / 2 \underline{\lambda}) \left (\| \bm{\hat{\beta}}_0 - \bm{\beta}^*_0\|^2 - \| \bm{\hat{\beta}}_T - \bm{\beta}^*_T\| ^2 \right)$.

Additionally, \cite[Lemma 2]{LesageLandryTaylorShames} yields
\begin{align}
    \| \bm{\hat{\beta}}_k - \bm{\beta}^*_k\| &= \| \bm{\hat{\beta}}_k - \bm{\beta}^*_k - \bm{\beta}^*_{k-1} + \bm{\beta}^*_{k-1}\| \nonumber \\
    & \leq  \| \bm{\hat{\beta}}_k  - \bm{\beta}^*_{k-1}\| + \| \bm{\beta}^*_{k-1} - \bm{\beta}^*_k\| \nonumber \\
    &\leq \dfrac{3\overline{k} L_g}{2\underline{\lambda}} \| \bm{\hat{\beta}}_{k-1}  - \bm{\beta}^*_{k-1}\|^2 + \| \bm{\beta}^*_{k-1} - \bm{\beta}^*_k\| \nonumber \\
    &\leq \dfrac{3\overline{k} L_g}{2\underline{\lambda}} \| \bm{\hat{\beta}}_{k-1}  - \bm{\beta}^*_{k-1}\|^2 + \bar{v}. \label{eqn:e_k_inequality}
\end{align}
If $\bar{v} \leq \underline{\lambda}/(6 \overline{k} L_g)$, then \eqref{eqn:e_k_inequality} gives
\begin{equation}
    \lim\sup_{k\rightarrow \infty} \| \bm{\hat{\beta}}_k - \bm{\beta}^*_k\| \leq  \dfrac{\underline{\lambda} }{3 \overline{k} L_g}.
\end{equation}

\subsection{Computationally efficient approximate ONM}
\label{sec:approximate_Newton}
The online Newton's method (ONM) \eqref{eqn:online_Newton_exact} requires computation of the gradient $\nabla J_k (\bm{\hat{\beta}}_k)$ and Hessian $\nabla^2 J_k (\bm{\hat{\beta}}_k)$, both of which are summations where the number of terms will increase with $k$. This can be computationally intensive for large $k$.
For more efficient numerical implementation, we use a recursive formulation to approximately capture the accumulated cost, and derive the approximate online Newton step for minimizing it. 
Under this approximation, the computational cost per iteration does not increase over time. Moreover, the computation required  is significantly less than the sequential Monte Carlo approach of \cite{LeongZamani_SP}, see Section~\ref{sec:numerical}.

Write the cost function as 
$$ J_k (\bm{\beta}) = J_{k-1} (\bm{\beta}) + g_k (\bm{\beta}). $$
First, motivated by Assumption~\ref{assumption_Lesage_Landry} (i), we assume that $\bm{\hat{\beta}}_k$ will approximately minimize $J_{k-1}(\cdot)$, so that by the first order condition
$\nabla J_{k-1} (\bm{\hat{\beta}}_k) \approx 0$, and hence $ \nabla J_k (\bm{\hat{\beta}}_k) \approx \nabla g_k (\bm{\hat{\beta}}_k) $.

Next, the Hessian of $J_k(\cdot)$ at $\hat{\bm{\beta}}_{k}$ is approximated as
\begin{equation}\label{eq:recursive_hess}
\begin{split}
\nabla^2 J_k(\hat{\bm{\beta}}_{k})  & = \nabla^2 J_{k-1}(\hat{\bm{\beta}}_{k}) + \nabla^2 g_k(\hat{\bm{\beta}}_{k}) \\
& \approx \nabla^2 J_{k-1}(\hat{\bm{\beta}}_{k-1}) + \nabla^2 g_k(\hat{\bm{\beta}}_{k}),
\end{split}
\end{equation}
where in the second line of \eqref{eq:recursive_hess}, motivated by Assumption~\ref{assumption_Lesage_Landry} (iii), we make the assumption that 
$\nabla^2 J_{k-1}(\hat{\bm{\beta}}_{k}) \approx \nabla^2 J_{k-1}(\hat{\bm{\beta}}_{k-1})$.
This allows us to use the Hessian $ \nabla^2 J_{k-1}(\hat{\bm{\beta}}_{k-1})$ from the previous time step, without having to recalculate the terms $g_t(\hat{\bm{\beta}}_{k}), t=0,\dots,k-1$ based on the new estimate $\hat{\bm{\beta}}_{k}$. 

We can then consider the iterations for an approximate online Newton's method of the form 
\begin{equation}
\label{eqn:online_Newton_approx}
  \bm{\hat{\beta}}_{k+1} = \bm{\hat{\beta}}_k - \left( H_k(\bm{\hat{\beta}}_k) \right)^{-1} G_k (\bm{\hat{\beta}}_k),
\end{equation}
where  
\begin{equation}
\label{eqn:J_tilde}
\begin{split}
G_k (\bm{\hat{\beta}}_k) & = \nabla g_k (\bm{\hat{\beta}}_k) \\
H_k(\hat{\bm{\beta}}_{k}) & =  H_{k-1}(\hat{\bm{\beta}}_{k-1}) + \nabla^2 g_k(\hat{\bm{\beta}}_{k}).
\end{split}
\end{equation} 

We next provide a numerically robust formulation of the approximate Newton iteration~\eqref{eqn:online_Newton_approx} and~\eqref{eqn:J_tilde}. The first idea is to use a full rank initialization $(H_0(\bm{\hat{\beta}}_{0}))^{-1}=\epsilon I$ for some $\epsilon > 0$.
Recall the Hessian of the per stage cost
$$ \nabla^2 g_k (\bm{\hat{\beta}}_{k}) = \frac{ \eta^2 \tilde{z}_{k}^2 \exp(\eta \tilde{z}_{k}(\bm{\hat{\beta}}_{k}^T \mathbf{K}(\mathbf{x}_{k}) -  \tau)) }{\big(1+\exp(\eta \tilde{z}_{k}(\bm{\hat{\beta}}_{k}^T \mathbf{K}(\mathbf{x}_{k}) -  \tau)) \big)^2}  \mathbf{K}(\mathbf{x}_{k}) \mathbf{K}^T(\mathbf{x}_{k}).$$
In short this can be written as 
$$
\nabla^2 g_k (\bm{\hat{\beta}}_{k}) = h(\mathbf{x}_{k},\bm{\hat{\beta}}_{k})\mathbf{K}(\mathbf{x}_{k}) \mathbf{K}^T(\mathbf{x}_{k}),
$$
where $h(\mathbf{x}_{k},\bm{\hat{\beta}}_{k})\in\mathbb{R}$. 
Then the rank one update (see e.g. \cite{CalafioreElGhaoui,HornJohnson}) can be used to update the next and subsequent iterations:
\begin{align*}
        & (H_{k}(\bm{\hat{\beta}}_{k}))^{-1}  =  (H_{k-1}(\bm{\hat{\beta}}_{k-1}))^{-1} \\ & \quad \times \Big(I-\dfrac{ h(\mathbf{x}_{k},\bm{\hat{\beta}}_{k})\mathbf{K}(\mathbf{x}_{k}) \mathbf{K}^T(\mathbf{x}_{k}) (H_{k-1}(\bm{\hat{\beta}}_{k-1}))^{-1} }{1 + h(\mathbf{x}_{k},\bm{\hat{\beta}}_{k}) \mathbf{K}^T(\mathbf{x}_{k})(H_{k-1}(\bm{\hat{\beta}}_{k-1}))^{-1}\mathbf{K}(\mathbf{x}_{k})}\Big).
\end{align*}
\begin{remark}
The formal treatment of the question of robustness of \eqref{eqn:J_tilde} is beyond the scope of this letter. However, in light of \cite{colabufo2020newton} and the numerical example presented later in this letter, we conjecture that the proposed approximate ONM is robust to the introduced approximation errors.
\end{remark}
\section{Active Sensing}
\label{sec:active_sensing}
In this section we consider the problem of adaptively selecting the next measurement location $\mathbf{x}_{k+1}$, based on the measurements $z_0,\dots,z_k$ collected so far, with the aim of achieving better estimation performance. We refer to this as \emph{active sensing} \cite{Kreucher_proceedings}. The active sensing mechanism of \cite{LeongZamani_SP} was based optimizing an information measure \cite{RisticMorelandeGunatilaka} which does not extend in a straightforward manner to the current approach. Based on Assumption \ref{assumption_Lesage_Landry} and informed by the results of Section \ref{sec:convergence}, having a larger $\underline{\lambda}$ may be desirable in terms of convergence behaviour. Motivated by this, in this letter we propose to choose position $\mathbf{x}_{k+1}$ to be the one that maximizes the minimum eigenvalue of the expected Hessian \eqref{eqn:expected_Hessian_exact}.

For a given future position $\mathbf{x}'$, consider an ``expected Hessian'' term:
\begin{equation}\small
\label{eqn:expected_Hessian_exact}
\begin{split}
& \textnormal{Hess}^+(\mathbf{x}') \triangleq \nabla^2 J_k(\bm{\hat{\beta}}_{k+1})  + \mathbb{P}(\tilde{z}_{k+1} = 1) \\ & \quad \quad \times \frac{ \eta^2 \exp(\eta (\bm{\hat{\beta}}_{k+1}^T \mathbf{K}(\mathbf{x}') -  \tau)) }{\big(1+\exp(\eta (\bm{\hat{\beta}}_{k+1}^T \mathbf{K}(\mathbf{x}') -  \tau)) \big)^2}  \mathbf{K}(\mathbf{x}') \mathbf{K}^T(\mathbf{x}') \\
 & \quad + \mathbb{P}(\tilde{z}_{k+1} = -1) \\ & \quad \quad \times \frac{ \eta^2 \exp(-\eta (\bm{\hat{\beta}}_{k+1}^T \mathbf{K}(\mathbf{x}') -  \tau)) }{\big(1+\exp(-\eta (\bm{\hat{\beta}}_{k+1}^T \mathbf{K}(\mathbf{x}') -  \tau)) \big)^2}  \mathbf{K}(\mathbf{x}') \mathbf{K}^T(\mathbf{x}') \\
 &= \nabla^2 J_k  (\bm{\hat{\beta}}_{k+1}) \\ & \quad +  \frac{ \eta^2 \exp(\eta (\bm{\hat{\beta}}_{k+1}^T \mathbf{K}(\mathbf{x}') -  \tau)) }{\big(1+\exp(\eta (\bm{\hat{\beta}}_{k+1}^T \mathbf{K}(\mathbf{x}') -  \tau)) \big)^2}  \mathbf{K}(\mathbf{x}') \mathbf{K}^T(\mathbf{x}') 
\end{split}
\end{equation}
where the last equality holds since $\exp(-y)/(1+\exp(-y))^2 = \exp(y)/(1+\exp(y))^2$ and $\mathbb{P}(\tilde{z}_{k+1} = 1) + \mathbb{P}(\tilde{z}_{k+1} = -1) = 1$. 


The optimization problem of interest, to maximize the minimum eigenvalue of the expected Hessian \eqref{eqn:expected_Hessian_exact}, is
$$\mathbf{x}_{k+1} = \textnormal{arg} \max\limits_{\mathbf{x}' \in \mathcal{X}_{k+1}} \lambda_{\min} (\textnormal{Hess}^+(\mathbf{x'})),$$
where $ \mathcal{X}_{k+1}$ is the set of possible future positions\footnote{The set $ \mathcal{X}_{k+1}$ may, e.g., capture the set of reachable positions from the current state of the mobile sensor platform.}.

In the case where some of the positions in $ \mathcal{X}_{k+1}$ may be far from the vehicle's current position, we may instead decide to only travel a limited distance $\rho$ towards the intended position before taking a measurement. That is, we collect a new measurement at position
$\mathbf{x}_{k+1} = \mathbf{x}_{k} + \rho \mathbf{d}_k$, where $\mathbf{d}_k = (\tilde{\mathbf{x}}_{k+1} - \mathbf{x}_{k})/\|\tilde{\mathbf{x}}_{k+1} - \mathbf{x}_{k}\|$ and
$\tilde{\mathbf{x}}_{k+1} = \textnormal{arg} \max\limits_{\mathbf{x}' \in \mathcal{X}_{k+1}} \lambda_{\min} (\textnormal{Hess}^+(\mathbf{x'})).$
One possibility for $ \mathcal{X}_{k+1}$ with this approach is then the set of centers of the basis functions, i.e. 
$ \mathcal{X}_{k+1} = \{\mathbf{c}_1, \dots, \mathbf{c}_p \}$. We can also smooth out the motion by taking a convex combination of the new ($\mathbf{d}_k$) and previous ($\tilde{\mathbf{d}}_{k-1}$) directions, namely 
$\mathbf{x}_{k+1} = \mathbf{x}_{k}  + \rho \tilde{\mathbf{d}}_k$, where
$\tilde{\mathbf{d}}_k = (\alpha \mathbf{d}_k + (1-\alpha) \tilde{\mathbf{d}}_{k-1})/\|\alpha \mathbf{d}_k + (1-\alpha) \tilde{\mathbf{d}}_{k-1} \|$, for some $\alpha \in [0,1]$. We have found this smoothing for values of $\alpha$ of around 0.3-0.5 to significantly improve performance.

For the approximate ONM of Section \ref{sec:approximate_Newton}, we can similarly choose positions to maximize the minimum eigenvalue of an expected Hessian term, but instead of \eqref{eqn:expected_Hessian_exact} we use instead
\begin{align*}
&\widetilde{\textnormal{Hess}}^+ (\mathbf{x}')   \triangleq H_k  (\bm{\hat{\beta}}_{k}) \\ & \quad +  \frac{ \eta^2 \exp(\eta (\bm{\hat{\beta}}_{k+1}^T \mathbf{K}(\mathbf{x}') -  \tau)) }{\big(1+\exp(\eta (\bm{\hat{\beta}}_{k+1}^T \mathbf{K}(\mathbf{x}') -  \tau)) \big)^2}  \mathbf{K}(\mathbf{x}') \mathbf{K}^T(\mathbf{x}').
\end{align*}

\section{Numerical Results}
\label{sec:numerical}

We consider a $[0,100\textnormal{m}] \times [0,100\textnormal{m}]$ area of interest $\mathcal{S}$. 
We consider randomly generated true fields using the model \eqref{field_model} with $p^\textnormal{true}=4$, with parameters $\beta_i^\textnormal{true} \sim U(0.7,1.4)$, each component of $\textbf{c}_i^\textnormal{true}$ $\sim U(5,95)$, and $\sigma_i^\textnormal{true} \sim U(25,45)$, for $i=1,\dots,p^\textnormal{true}$.
 In the measurement model \eqref{eqn:noisy_measurement}, we assume $v(\mathbf{x}) \sim \mathcal{N}(0,\sigma_v^2)$ is i.i.d. Gaussian noise, with $\sigma_v^2 = 0.1$, and the sensor threshold $\tau = 1$. 
\begin{figure}[t!]
\centering 
\begin{subfigure}[t]{0.241\textwidth}
\includegraphics[width=.9\textwidth]{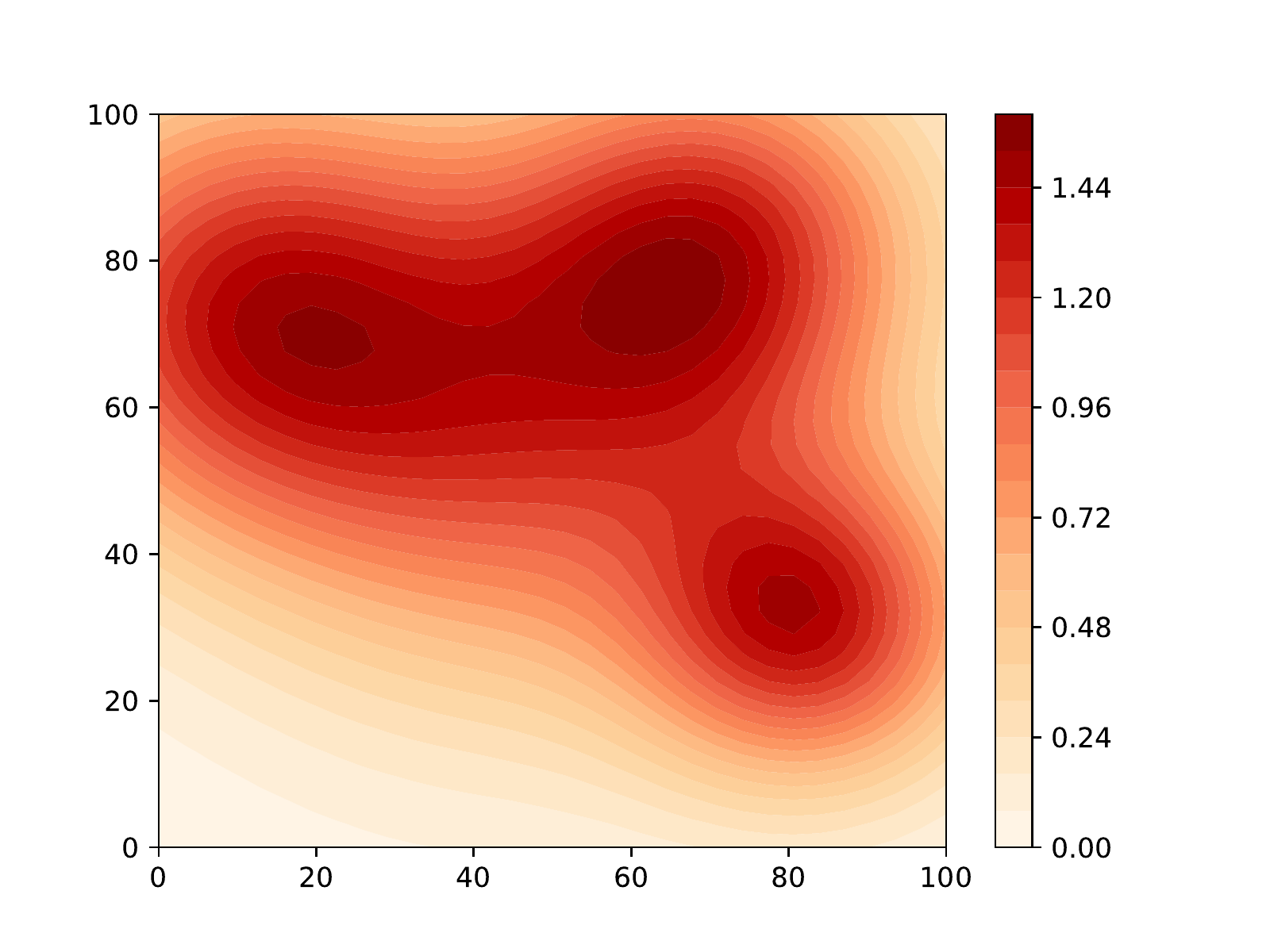} 
\end{subfigure} 
\begin{subfigure}[t]{0.241\textwidth}
\includegraphics[width=.9\textwidth]{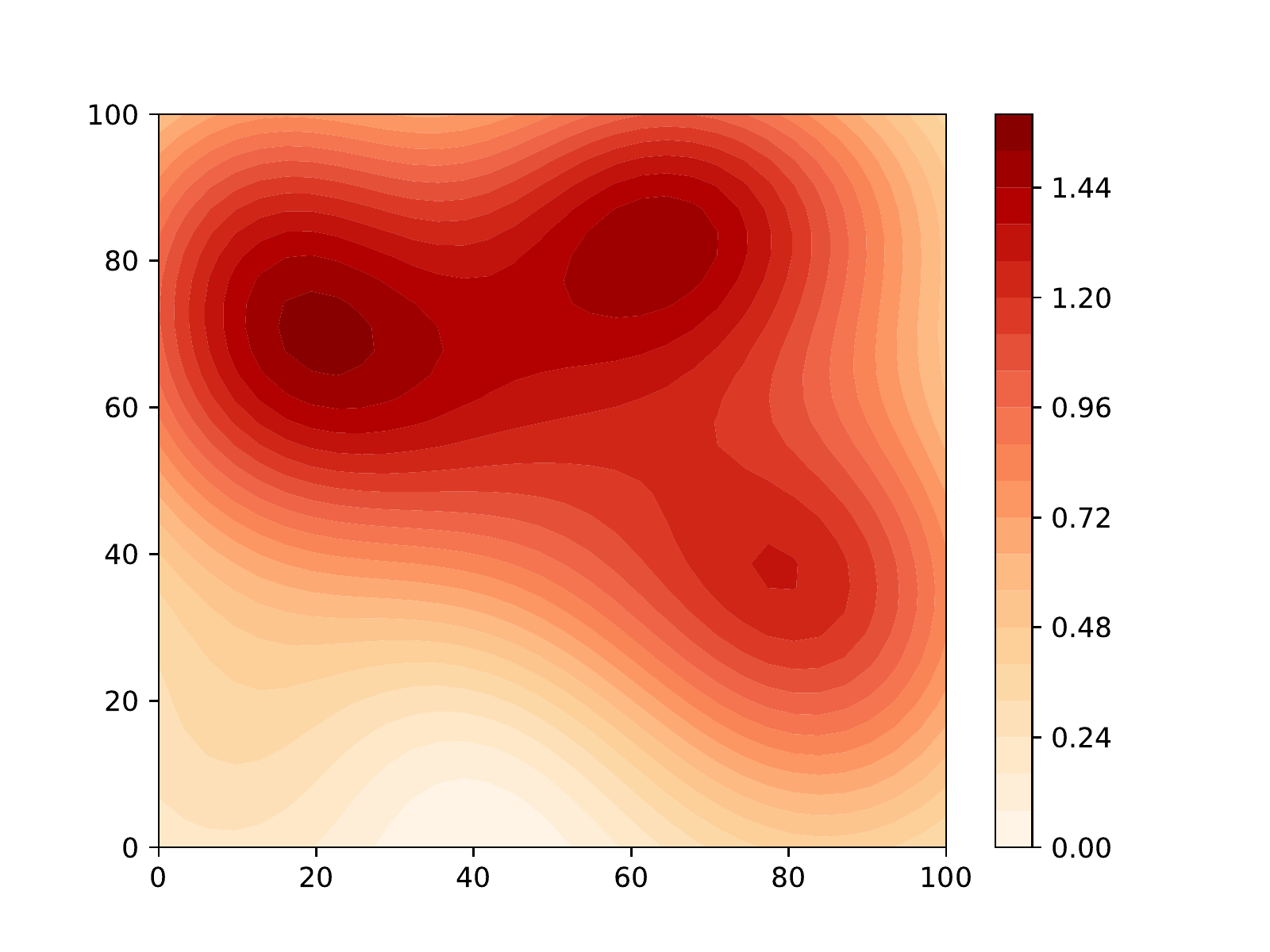} 
\end{subfigure} 
\caption{True field (left) and estimated field after 1000 measurements (right)}
\label{fig:field_plots}
\end{figure}


First, we use the approximate online Newton's method of Section \ref{sec:approximate_Newton} together with the active sensing method of Section \ref{sec:active_sensing}. The logistic function parameter is chosen as $\eta = 5$. We use $p=16$ basis functions, with centers $\mathbf{c}_i \in \{12.5, 37.5, 62.5, 87.5\}^2$ and $\sigma_i = 25, i=1,\dots,16$. The initial estimate $\bm{\hat{\beta}}_{0}$ is randomly chosen with each component $\sim U(0,1)$. The full rank initialization $(H_0(\bm{\hat{\beta}}_{0}))^{-1}= \epsilon I$ with $\epsilon=0.1$. 
In active sensing we use $ \mathcal{X}_{k+1}= \{ \mathbf{c}_1, \dots,\mathbf{c}_{16}$\},  $\rho=5\textnormal{m}$, and $\alpha=0.4$. 

A plot of a sample randomly generated field that we wish to estimate is given on the left of Fig.~\ref{fig:field_plots}. On the right of Fig.~\ref{fig:field_plots} we show the estimated field after a single simulation run consisting of the collection of 1000 measurements. We see that the approximate ONM can capture the qualitative shape of the true field. 

We next compare the performance of approximate ONM, ``exact'' ONM, and the sequential Monte Carlo (SMC) approach of \cite{LeongZamani_SP}. We note that for this particular example, during the initial steps, the exact ONM is not numerically stable due to the Hessian being close to singular (which violates Assumption~\ref{assumption_Lesage_Landry} (ii)). We used a hybrid Newton implementation that used a damping multiplier of $0.1$ for the derivative until the Hessian was large enough, at which point we switched to the original exact ONM. Moreover a small regularization factor of $0.1I$ was added to the Hessian when it got close to being singular. For performance evaluation, we use the following Mean Squared Error (MSE) type measure:
\begin{equation}
\label{MSE_field_probs}
\frac{1}{|\tilde{\mathcal{S}}|} \sum_{\mathbf{x} \in \tilde{\mathcal{S}}} \left( \mathbb{P} (z = 1 | \bm{\beta} ; \mathbf{x}) -  \mathbb{P} (z = 1 | \bm{\hat{\beta}}_{k} ; \mathbf{x})\right)^2,
\end{equation}
which was shown in \cite{LeongZamani_SP} to be a useful measure in comparing the closeness of two fields. In \eqref{MSE_field_probs} $\tilde{\mathcal{S}}$ is a discretization of $\mathcal{S}$ where the $x$ and $y$ axes are each discretized into 32 values, giving cardinality $|\tilde{\mathcal{S}}| = 32^2 = 1024$. From the model of Section \ref{sec:system_model}, we can derive the expression $\mathbb{P} (z = 1 | \bm{\beta} ; \mathbf{x})  = 1 -\Phi \Big(  \big(\tau - \sum_{i=1}^p \beta_i K_i (\mathbf{x})\big)/\sigma_v \Big)$, where $\Phi(\cdot)$ is the cdf of the standard normal distribution. 

\begin{figure}[t]
\centering 
\includegraphics[scale=0.5]{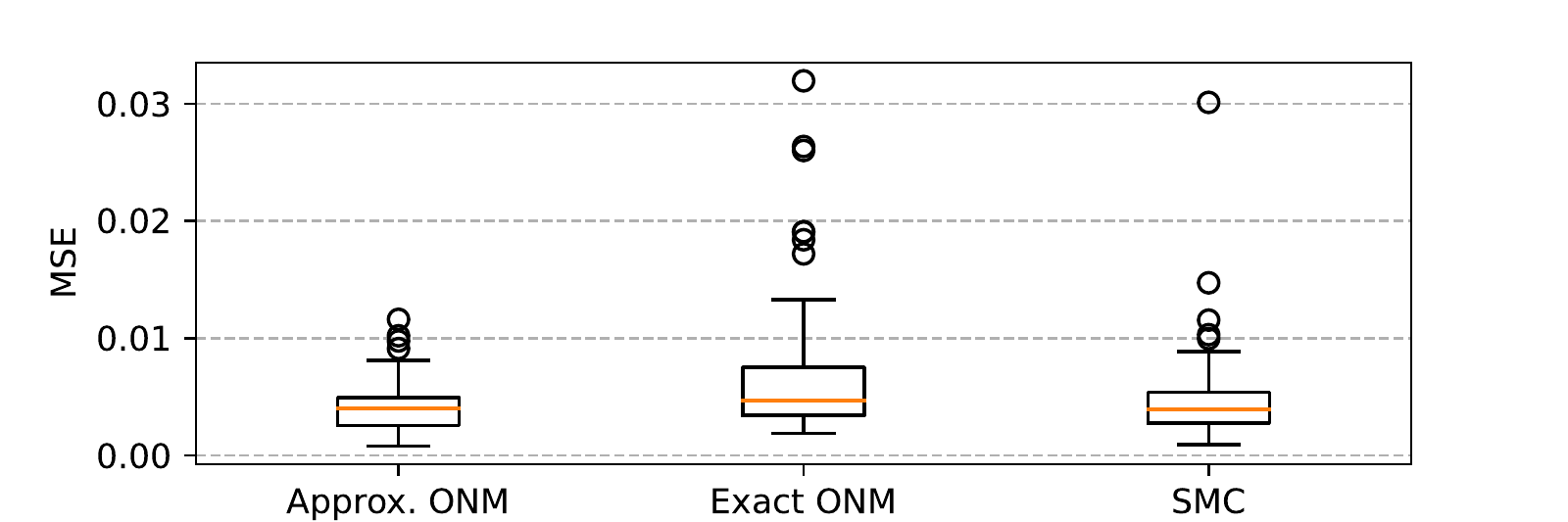} 
\caption{Box plots corresponding to the performance of different methods after 1000 measurements over 100 scenarios.}
\label{fig:boxplot_comparison}
\end{figure}

\begin{figure}[t!]
\centering 
\begin{subfigure}[t]{0.241\textwidth}
\includegraphics[width=\textwidth]{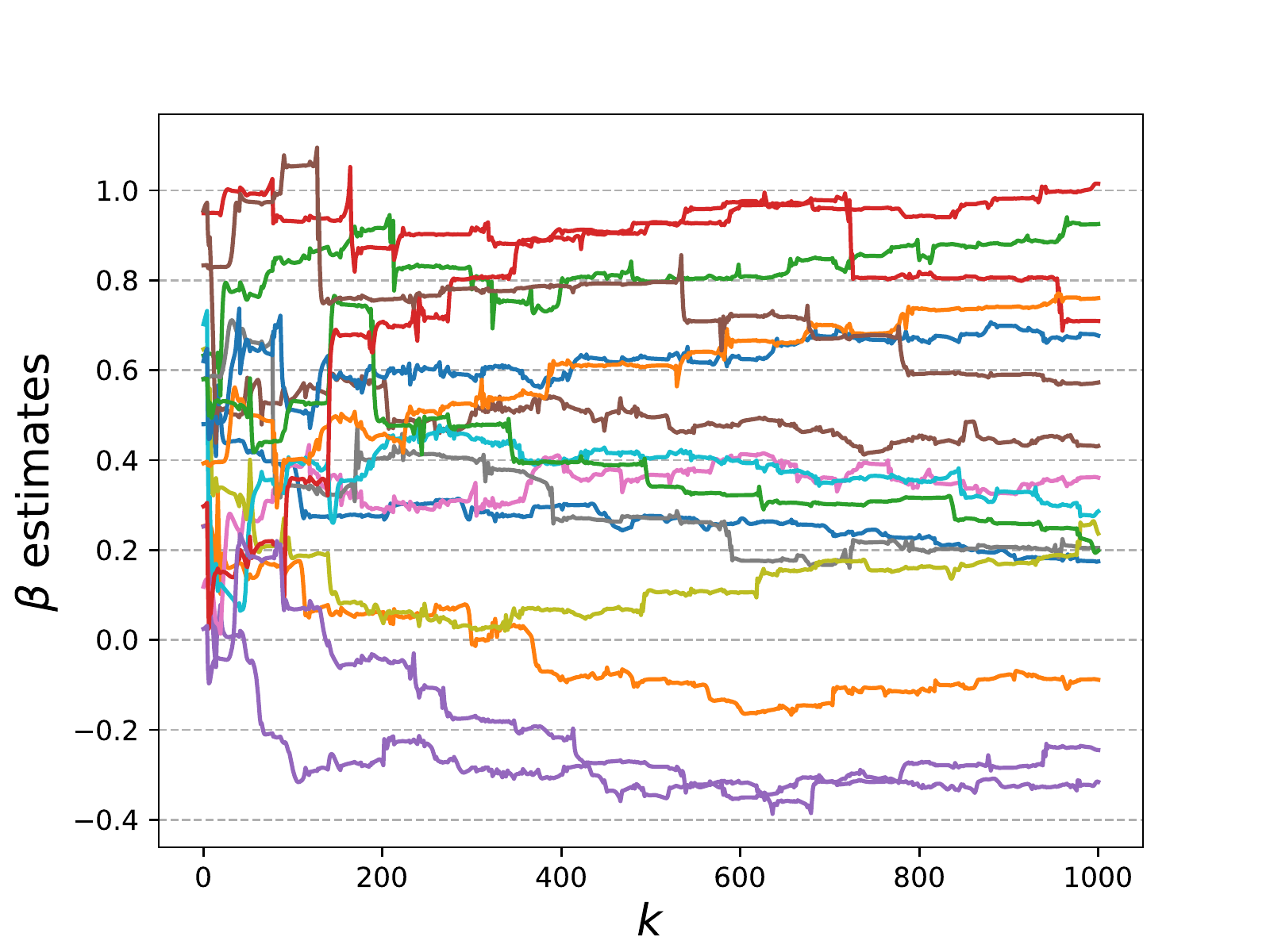} 
\end{subfigure} 
\begin{subfigure}[t]{0.241\textwidth}
\includegraphics[width=\textwidth]{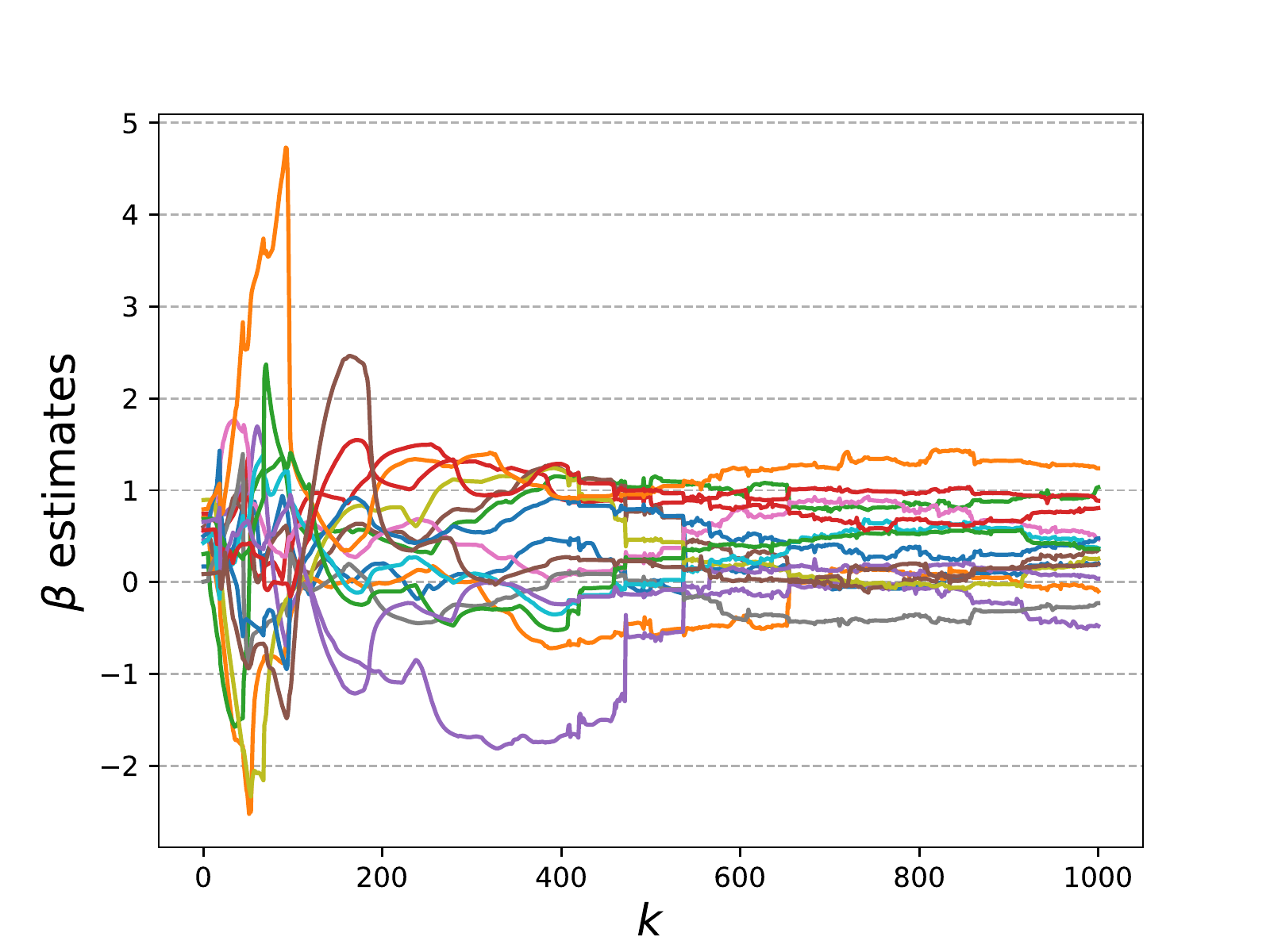} 
\end{subfigure} 
\caption{Parameter estimates for approximate ONM (left) and exact ONM (right)}
\label{fig:beta_hat_plot}
\end{figure}

In Fig. \ref{fig:boxplot_comparison} we show boxplots for the MSE's at $k=1000$, obtained from simulation runs for 100 different fields. The computations are done using a Core i7-9700 desktop with 16 Gb of RAM. The results are summarised in Table \ref{table:algorithm_comparison}. 
The approximate ONM and SMC methods generally perform similar to each other. However, we see that the approximate ONM performs more consistently than the exact ONM and SMC approaches, with fewer outliers that have large MSEs. Moreover, the running time for the approximate ONM is significantly faster than the other two methods. In regards to the poorer performance of the exact ONM compared to the approximate ONM, Fig. \ref{fig:beta_hat_plot} plots the estimates of $\bm{\beta}$ over time for a single run of both the approximate and exact ONMs. We see that the parameter estimates for the exact ONM have larger variations than the approximate ONM, especially in the earlier iterations. This might be due the fact that the exact ONM is sensitive to initial condition and as a result had to be modified in our implementation. A rigorous study of this phenomenon and the dependency of the methods to the initial condition is a potential future research direction.

\begin{table}[t!]
\caption{Performance of different methods}
\centering
\begin{tabular}{|c|c|c|c|c|} \hline
Method & median MSE  & min MSE & max MSE & Time/Run (s)  \\ \hline \hline 
Approx.~ONM & \textbf{0.00374} & 0.00108 & \textbf{0.0121} & \textbf{1.3} \\ \hline
Exact ONM & 0.00470 & 0.00187 & 0.0319 & 13.0 \\ \hline
SMC \cite{LeongZamani_SP} & 0.00391 & \textbf{0.00092} & 0.0301 & 9.5 \\ \hline
\end{tabular}
\label{table:algorithm_comparison}
\end{table}

\section{Conclusion}
An approach based on online logistic regression for the problem of field estimation using binary measurements has been proposed. The developed algorithm is less computationally intensive than an existing sequential Monte Carlo approach, while having more reliable estimation performance. 

\bibliography{IEEEabrv,source_localization}
\bibliographystyle{IEEEtran} 

\end{document}